\documentclass[12pt]{article}
\usepackage{latexsym}
\usepackage{amsmath}
\usepackage{amsfonts}
\usepackage{amssymb}
\usepackage{graphicx}
\usepackage{slashbox}
 
\newcommand{\mR}{{\mathbb R}}
\newcommand{\mH}{{\mathcal H}}

\title{Nonrelativistic conformal groups and their dynamical realizations}
\author{K. Andrzejewski\thanks{e-mail: k-andrzejewski@uni.lodz.pl},
J. Gonera, P. Ma\'slanka\\
\small Department of Theoretical Physics and Computer Science, \\
\small University of \L\'od\'z,\\
\small Pomorska 149/153, 90-236 {\L}\'od\'z, Poland
}
\date{}
\begin{document}
\maketitle
\begin{abstract}
Nonrelativistic conformal groups, indexed by $l=\frac{N}{2}$,  are analyzed. Under the assumption that the "mass" parametrizing  the central extension is nonvanishing the coadjoint orbits are classified and described in terms of convenient variables. It is shown that the corresponding dynamical system describes, within Ostrogradski framework, the nonrelativistic particle obeying $(N+1)$-th order equation of motion. As a special case, the Schr\"odinger group      and the standard Newton equations are obtained for $N=1$ ($l=\frac{1}{2}$).
\end{abstract}
\section{Introduction}
Historically, the structure which is now called the Schr\"odinger group has been discovered in XIX century in the context of classical mechanics \cite{b1} and heat equation \cite{b2}. It has been rediscovered in XX century as the maximal symmetry group of free motion in quantum mechanics \cite{b3}-\cite{b10}. Much attention has been paid to the structure of Schr\"odinger group and its geometrical status \cite{b7}, \cite{b11}-\cite{b15}.
\par
The Schr\"odinger group, when supplemented with space dilatation transformations becomes $l=\frac{1}{2}$ member of the whole family of nonrelativistic conformal groups \cite{b16,b17}, indexed by halfinteger $l$. Various structural, geometric and physical aspects of the resulting Lie algebras have been intensively studied \cite{b18}-\cite{b32e}. For $l=\frac{N}{2}$, $N$-odd ($N$-even in the case of dimension two), the nonrelativistic conformal algebra admits central extension. Then, as it has been shown in Ref. \cite{b29}, it becomes the symmetry algebra of free nonrelativistic particle obeying $(N+1)$-th order equation of motion.
\par
 In the present paper we use the orbit method \cite{b33}-\cite{b36} to construct the most general dynamical systems on which the nonrelativistic conformal groups act transitively as symmetries. We find that the basic variables are coordinates and momenta together with "internal" variables obeying $SU(2)$ commutation rules (in the sense of Poisson brackets) and underlying trivial dynamics; the remaining  internal variables obey $SL(2,\mR)$ (or $SO(2,1)$) commutation rules and equation of motion of conformal quantum mechanics \cite{b37} in global formulation \cite{b38}.
\par
All symmetry generators split into two parts: the external one constructed out of coordinates and momenta  (like orbital angular momentum) and internal one (like spin). The symmetry transformations are implemented as canonical transformations. 
\par
The standard free dynamics is obtained by selecting the trivial orbit for $SL(2,\mR)$ variables. 
\par
The results heavily rely on the fact that the conformal algebras under considerations admit central extensions. For   vanishing "mass" parameters (as well as for conformal algebras which do not admit central extension)  the classification of orbits is more complicated and the physical interpretation in such cases remains slightly obscure.
\section{The Schr\"odinger symmetry}   
We start with the $l=\frac{1}{2}$ Galilean conformal algebra (according to the terminology of Ref. \cite{b16,b17}). It consists of rotations $\vec J$, translations $\vec{P}$, boosts $\vec{B}$ and time translations $H$ which form the Galilean algebra, together with dilatations $D$, conformal transformations $K$ and, finally, space dilatations $D_s$. The nontrivial commutation rules read
\begin{align}
\label{e1}
&[J_i,J_k]=i\epsilon_{ikl}J_l,\quad [J_i,P_k]=i\epsilon_{ikl}P_l,\quad [J_i,B_k]=i\epsilon_{ikl}B_l,\nonumber\\
&[B_i,H]=iP_i,\nonumber\\
&[D,H]=iH,\quad [D,K]=-iK, \quad[K,H]=2iD,\\
&[D,P_i]=\frac{i}{2}P_i,\quad [D,B_i]=\frac{-i}{2}B_i,\quad [K,P_i]=iB_i,\nonumber\\
&[D_s,P_i]=iP_i,\quad [D_s,B_i]=iB_i.\nonumber
\end{align}
Deleting $D_s$ one obtains twelvedimensional  Schr\"odinger algebra which admits, similarly to the Galilei algebra,  central extension defined by additional nontrivial commutator
\begin{equation}
\label{e2}
[B_i,P_k]=iM\delta_{ik}.
\end{equation}
The structure of centrally extended Schr\"odinger algebra is well known. First, we have $su(2)$ (or $so(3)$) algebra spanned by $J_i'$s; furthermore, $H$, $D$ and $K$ span the conformal algebra which is isomorphic to $so(2,1)$ (or $sl(2,\mR)$). To see this one defines 
\begin{equation}
\label{e3}
N^0=\frac{1}{2}(H+K), \quad N^1=\frac{1}{2}(K-H),\quad N^2=D,
\end{equation}
which yields 
\begin{equation}
\label{e4}
[N^\alpha,N^\beta]=i{\epsilon^{\alpha\beta}}_\gamma N^\gamma,\quad 
\alpha,\beta,\gamma=0,1,2;
\end{equation}
where $\epsilon^{012}=\epsilon_{012}=1$, and $g_{\mu\nu}=diag(+,-,-)$. Therefore $\vec{ J},H,K$ and $D$ span direct sum $su(2)\oplus so(2,1)$. Finally, $\vec{P},\vec{B}$ and $M$ form a nilpotent algebra which, at the same time, carries a representation of $su(2)\oplus so(2,1)$. To express this fact in compact way we define the spinor representation of $so(2,1)$:
\begin{equation}
\label{e5}
\tilde N^0=\frac{1}{2}\sigma_2, \quad \tilde N^1=\frac{i}{2}\sigma_1,\quad \tilde N^2=\frac{i}{2}\sigma_3.
\end{equation}
Moreover, denoting $X_{1i}=P_i$, $X_{2i}=B_i$ one finds simple form of the action of $su(2)\oplus  so(2,1)$ on the space spanned by $\vec{P},\vec{B}$ and $M$ 
\begin{equation}
\label{e6}
[J_i,X_{ak}]=i\epsilon_{ikl}X_{al},\quad [N^\alpha,X_{ai}]=X_{bi}({\tilde N}^\alpha)_{ba},
\end{equation}
where $a,b=1,2$. The commutation rule (\ref{e2}) takes the form 
\begin{equation}
\label{e7}
[X_{ai},X_{bj}]=-iM\epsilon_{ab}\delta_{ij}.
\end{equation}    
The matrices $\tilde N^\alpha$ are all purely imaginary and span the defining representation of $sl(2,\mR)$. In fact, the group $SL(2,\mR)$ is nothing but the  group $Spin(2,1)^+$. The Schr\"odinger algebra can be thus integrated to the group $S=(SU(2)\times SL(2,\mR))\ltimes R_7$, where $R_7$ is sevendimensional nilpotent  group (topologically isomorphic to $\mR^7$) and the semidirect product is defined by the $D^{(1,\frac{1}{2})}\oplus D^{(0,0)}$ representation of $SU(2)\times SL(2,\mR)$.
\par
Let us consider the coadjoint action of Schr\"odinger group $S$. Denote the dual basis elements by $ \tilde{\vec{J}},\tilde{\vec{P}},\tilde{\vec{B}}$ etc. The general element of the dual space to the Lie algebra of $S$ is written as 
\begin{equation}
\label{e8}
X=\vec j\tilde{\vec J}+\vec \xi\tilde{\vec P}+ \vec\zeta\tilde{\vec B}+h\tilde H+d\tilde D+k\tilde K+m\tilde M.
\end{equation}
Having characterized the global structure of $S$ we could consider the full action of $S$ on $X$. However, for our purposes it is sufficient to compute the coadjoint action of one-parameter subgroups generated by the basic elements of the Lie algebra. The results are summarized in Table 1 below.
%%%%TABLE%%%%%%%%%
%\begin{center}
\begin{table}[!hb]
\caption{Coadjoint action of $S$.}
\centering
\scalebox{0.8}{
\begin{tabular}{|c|c|c|c|c|c|c|}\hline
\backslashbox{$Ad^*_g$}{$g$}& $e^{i\vec a\vec P}$&$e^{i\vec v\vec B}$&$e^{-i\tau H}$&$e^{i\lambda D}$&$e^{iuK}$&$e^{i\vec \omega\vec J}$\\ \hline
$\vec j'$&$\vec j-\vec a\times\vec \xi $& $\vec j-\vec v\times\vec \zeta$&$\vec j$&$\vec j$&$\vec j$&$\overrightarrow{Rj}$\\ \hline
$\vec \xi'$&$\vec \xi $&$\vec \xi+m\vec v$&$\vec \xi$&$e^{\frac \lambda2}\vec \xi$&$\vec \xi+u\vec\zeta$&$\overrightarrow{R\xi}$ \\ \hline
$\vec \zeta'$&$\vec \zeta-m\vec a $&$\vec \zeta$&$\vec \zeta+\tau \vec \xi$&$e^{-\frac \lambda2}\vec \zeta$&$\vec\zeta$&$\overrightarrow{R\zeta }$\\ \hline
$h'$&$h $&$h+\frac{m\vec v^2}{2}+\vec v\vec\xi$&$h$&$e^\lambda h$&$h+2ud+u^2k$&$h$\\ \hline
$d'$&$d-\frac 12\vec a\vec\xi $&$d+\frac 12 \vec v\vec\zeta$&$d+\tau h $&$d$&$d+uk$&$d$\\ \hline
$k'$&$k-\vec a\vec\zeta+\frac 12m\vec a^2 $&$k$&$k+2\tau d+\tau^2h$&$e^{-\lambda}k$&$k$&$k$\\ \hline
$m'$&$m $&$m$&$m$&$m$&$m$&$m$\\ \hline
\end{tabular}
}
\end{table}
\newline
here $(\overrightarrow{Rj})_k=R_{kl}j_l$ etc.
%\end{center}
%%%%%%%%%%%%%
\par In order to find the structure of coadjoint orbits note that $m$ is invariant under the coadjoint action of $S$. In what follows we assume that $m>0$ (in fact, it is sufficient to take $m\neq 0$). Once this assumption is made, the classification of orbits become quite simple. Using the results collected in Table 1 we conclude that each orbit contains the point corresponding to $\vec\xi=0,\vec\zeta=0$. Moreover, the stability subgroup of the submanifold $\vec\xi=0,\vec\zeta=0$ is $SU(2)\times SL(2,\mR)\times \mR$ where the last factor is the subgroup generated by $M$ and can be neglected. The orbits of $SU(2)\times SL(2,\mR)$ are the products of orbits of both factors. For $SU(2)$ any coadjoint orbit is a $2$-sphere (or a point) which can be parametrized by vector $\vec s$ of fixed length, $\vec s^2=s^2$.
To describe the orbits of $SL(2,\mR)$ (which is equivalent, as far as coadjoint action is concerned, to $SO(2,1)$) we define, in analogy with eq. (\ref{e3}), 
\begin{equation}
\label{e9}
\chi^0=\frac{1}{2}(h+k),\quad \chi^1= \frac{1}{2}(-h+k),\quad \chi^2=d.
\end{equation} 
Then, by standard arguments, the full list of orbits reads:
\begin{equation}
\label{e10}
\begin{split}
\mH_\sigma^+&=\{\chi^\mu:g_{\mu\nu}\chi^\mu\chi^\nu=\sigma^2, \chi^0>0\},\\
\mH_\sigma^-&=\{\chi^\mu:g_{\mu\nu}\chi^\mu\chi^\nu=\sigma^2, \chi^0<0\},\\
\mH_0^+&=\{\chi^\mu:g_{\mu\nu}\chi^\mu\chi^\nu=0, \chi^0>0\},\\
\mH_0^-&=\{\chi^\mu:g_{\mu\nu}\chi^\mu\chi^\nu=0, \chi^0<0\},\\
\mH_\sigma&=\{\chi^\mu:g_{\mu\nu}\chi^\mu\chi^\nu=-\sigma^2\},\\
\mH_0&=\{0\}.
\end{split}
\end{equation}
Consequently, any coadjoint orbit of $S$ (with nonvanishing $m$) contains the point
\begin{equation}
\label{e11}
\vec s\tilde{\vec J}+(\chi^0-\chi^1)\tilde H+\chi^2\tilde D+(\chi^0+\chi^1)\tilde K+m\tilde M,
\end{equation}
where $\vec s\in S^2$ and $\chi^\mu$ is a point on one of the manifolds $\mH$ listed above.
We see that any orbit is characterized by the values of $m,\vec s^2$, $\chi^2$  and, for $\chi^2\geq 0$, the sign of $\chi^0$.
 Let us note that the above invariants correspond to the Casimir operators of Schr\"odinger algebra 
\begin{equation}
\label{e12}
\begin{split}
C_1&=M,\\
C_2&=(M\vec J-\vec B\times\vec P)^2,\\
C_3&=\left(MH-\frac{\vec P^2}{2}\right)\left(MK-\frac{\vec B^2}{2}\right)+\left(MK-\frac{\vec B^2}{2}\right)\left(MH-\frac{\vec P^2}{2}\right)+\\
&-2\left(MD-\frac{\vec B\vec P}{4}-\frac{\vec P\vec B}{4}\right)^2.
\end{split}
\end{equation}
The whole coadjoint orbit of $S$ can be obtained by applying $g(\vec a)$ and $g(\vec v)$ to all points (\ref{e11}) with $\vec s$ and $\chi^\mu$  varying over their orbits. Calling $\vec a=-\vec x$ and $\vec v=\vec p/m$ one finds the following parametrization of coadjoint orbits
\begin{equation}
\label{e13}
\begin{split}
\vec j&=\vec x\times\vec p+\vec s,\\
\vec \xi&=\vec p,\\
\vec\zeta&=m\vec x,\\
h&=\frac{\vec p^2}{2m}+\chi^0-\chi^1,\\
d&=\frac{1}{2}\vec x\vec p+\chi^2,\\
k&=\frac{m}{2}\vec x^2+\chi^0+\chi^1.
\end{split}
\end{equation}
We see that the phase-space variables are $\vec x,\vec p, \vec s$ and $\chi^\mu$. The  Poisson brackets implied by Kirillov symplectic structure read
\begin{align}
\label{e14}
&\{x_i,p_k\}=\delta_{ik},\nonumber\\
&\{s_i,s_k\}=\epsilon_{ikl}s_l,\\
&\{\chi^\alpha,\chi^\beta\}={\epsilon^{\alpha\beta}}_\gamma\chi^\gamma\nonumber,
\end{align}
while the corresponding equations of motion take the form 
\begin{equation}
\label{e14b}
\begin{split}
&\dot{\vec x}=\frac{\vec p}{m},\quad \dot{\vec p}=0,\quad \dot{\vec s}=0,\\
&\dot\chi^0=\chi^2,\quad \dot\chi^1=\chi^2,\quad \dot\chi^2=-\chi^1+\chi^0.
\end{split}
\end{equation}
We can summarize our findings. The tendimensional orbits are parametrized by $\vec x,\vec p,\vec s$ and $\chi ^\mu$ subject to the constraints $\vec s^2=const.$ and $g_{\mu\nu}\chi^\mu\chi^\nu=const.$ and equipped with the symplectic structure defined by eqs. (\ref{e14}) and dynamics given by eqs. (\ref{e14b}).
\par
One can say that, besides the standard canonical variables $\vec x$ and $\vec p$, there are two kinds of "internal" degrees of freedom -- ordinary spin variables $\vec s$ and $SO(2,1)$ "spin" degrees of freedom $\chi^\mu$. Note that, contrary to the true spin variables, $\chi^\mu$ have nontrivial dynamics.
%%%%%%%%%%%%%%%%%%%
%%%%%%%%%%%%%%%%%%
\section{Special cases}
Making the trivial choice $\mH_0=\{0\}$ of the $SL(2,\mR)$ orbit one finds the standard realization of Schr\"odinger group as the symmetry of free dynamics. The structure of the phase space is the same as in the case of Galilei group except that  the internal energy (the Casimir of Galilei group) vanishes. The additional generators $K$ and $D$  are constructed as the elements of  enveloping algebra of Galilei algebra.
\par
The Schr\"odinger algebra contains also Newton-Hooke algebra as subalgebra. This is easily seen by redefining the Hamiltonian: $H\rightarrow H\pm\omega^2K$. The Galilei and Newton-Hooke algebras are not isomorphic. However, due to the fact that, in the special case under consideration, $K$ belongs to the  enveloping algebra of Galilei one, Newton-Hooke algebra is contained in this  enveloping algebra and reverse.
\par 
In the general case of arbitrary orbit of $SL(2,\mR)$ both Galilei and Newton-Hooke algebras/groups do not act transitively. However, one can reduce the phase space by abandoning the variables $\chi^\mu$ except the combination $\chi^0-\chi^1$ ($(1+\omega^2)\chi^0+(-1+\omega^2)\chi^1$) which is now viewed as a constant representing the value of internal energy for Galilei (Newton-Hooke) algebra. The reduced phase space coincides with the one obtained by applying the orbit method directly to the Galilei or Newton-Hooke groups. 
\section{Canonical transformations}
From the basic functions (\ref{e13}) one can construct the generators (in the sense of canonical formalism) of group transformations. Due to the fact that the Hamiltonian is an element of the Lie algebra of symmetry group the symmetry generators depend, in general, explicitly on time. To construct the explicitly time dependent genrators of symmetry transformation one notes that the dynamics induces an internal automorphism of Lie algebra of Schr\"odinger group. Therefore, the relevant generators (providing the integrals of motion which existence is implied by the symmetry under consideration) are obtained by inverting this automorphism. The result reads
\begin{align}
\label{e15}
&j_k=j_k(t),\quad p_k=p_k(t),\nonumber\\
&x_k=x_k(t)-\frac{t}{m}p_k(t),\quad h=h(t),\\
&k=k(t)-2td(t)+t^2h(t),\quad d=d(t)-th(t).\nonumber
\end{align}         
In order to find the transformation generated by left-hand sides of eqs. (\ref{e15}) let us note that the one-parameter group of symmetry transformations of canonical variables $\eta$
\begin{equation}
\label{e16}
\begin{split}
t'&=g_1(t;c)\simeq t+\delta c\tilde g_1(t),\\
\eta'(t')&=g_2(\eta(t),t;c)\simeq \eta(t)+\delta c \tilde g_2(\eta(t),t),
\end{split}
\end{equation}
is related to its canonical generator $G(t)$ via
\begin{equation}
\label{e17}
\delta_0\eta=\delta c\{\eta,G\},
\end{equation}
where 
\begin{equation}
\label{e18}
\delta_0\eta=\eta'(t)-\eta(t)=\delta c(\tilde g_2(\eta(t),t)-\dot \eta(t)\tilde g_1(t)).
\end{equation}
As an example consider the transformation generated by $k$. By comparing eq. (\ref{e15}) for $k$ and eq. (\ref{e18}) we find
\begin{equation}
\label{e19}
\tilde g_1(t)=-t^2.
\end{equation}
Integration of  eq. (\ref{e19}) gives
\begin{equation}
\label{e20}
t'=\frac{t}{1+ct}.
\end{equation}
Having described the transformation properties of time variable one determines that of $x_i$ and $p_i$. To this end it is convenient to use the simplified form of $k$ , $k_s=k(t)-2td(t)$ together with the replacement $t\rightarrow t/(1+ct)$:
\begin{equation}
\label{e21}
\frac{dx_i}{dc}=\{x_i,k-\frac{2t}{1+ct}d\}=-\frac{t}{1+ct}x_i,
\end{equation}
yielding 
\begin{equation}
\label{e22}
x_i'=\frac{x_i}{1+ct}.
\end{equation}
Analogusly
\begin{equation}
\label{e23}
p_i'=p_i(1+ct)-mcx_i.
\end{equation}
Similarly, one can consider the action of conformal transformation on "internal" variables.
\par
The action of conformal transformation on time variable, eq. (\ref{e20}), can be extended to the whole Schr\"odinger group. In fact, by deleting the Hamiltonian $H$ one obtains the subgroup of   $S$. Therefore, it is possible to define the nonlinear action of Schr\"odinger group on onedimensional coset space. It is singular (cf. eq. (\ref{e20})) if one uses exponential parametrization because the latter provides only local map. Taking into account global topology requires more care \cite{b24}. The action of other generators may be described in a similar way.
\section{N-Galilean Conformal Symmetry}
Higher dimensional nonrelativistic conformal algebras are constructed according to the following unique scheme. One takes the direct sum $su(2)\oplus sl(2,\mR)\oplus\mR$, where the last term corresponds to the spatial dilatation $D_s$. This is supplemented by $3(N+1)$ Abelian algebra (here $l=N/2$) which carries the $D^{(1,\frac{N}{2})}$ representation of $SU(2)\otimes SL(2,\mR)$; moreover, all new generators correspond to the eigenvalue $1$ of $D_s$. Call $\vec C_i=(C^a_i, \quad a=1,2,3)$, $
i=0,1,\ldots,N$, the new generators. The relevant commutation rules involving $\vec C_i$ read
\begin{align}
\label{e24}
[D_s,C_j^a]&=iC_j^a,\nonumber\\
[J^a,C_j^b]&=i\epsilon_{abd}C_j^d,\nonumber\\
[H,C_j^a]&=-ijC_{j-1}^a,\\
[D,C_j^a]&=i(\frac N2-j)C_j^a,\nonumber\\
[K,C_j^a]&=i(N-j)C_{j+1}^a.\nonumber
\end{align}
As previously we delete the space dilatation operator $D_s$ and consider the question of the existence of central extension of the Abelian algebra spanned by $\vec C's$. To solve  it one can consider the relevant Jacobi identities or analyze the transformation properties under $SU(2)\times SL(2,\mR)$. The second order $SU(2)$ invariant  tensor, i.e. Kronecker delta $\delta^{ab}$ in arbitrary dimension (and  tensor $\epsilon^{ab}$ for dimension two),  is symmetric (antisymmetric, respectively), so the existence of central extension is equivalent to the existence antisymmetric  (symmetric) $SL(2,\mR)$ invariant tensor. Taking into account that $N+1$-dimensional irreducible representations of $SL(2,\mR)$ may be obtained from symmetrized  tensor product of $N$ basic representation one easily concludes that an invariant antisymmetric (symmetric) tensor exists only for $N$ odd (for $N$ even  in the case dimension two)  (see Ref. \cite{b14}).
\subsection{N-odd}
In this case the relevant central extension reads \cite{b29}
\begin{equation}
\label{e25}
[C_j^a,C_k^b]=i\delta^{ab}\delta^{N,j+k}(-1)^{\frac{k-j+1}{2}}k!j!M,
\end{equation}
for $j,k=0,1,\ldots,N$ and $a,b=1,2,3$. In order to classify the coadjoint orbits we put, in analogy to eq. (\ref{e8}),
\begin{equation}
\label{e26}
X=\vec j\tilde{\vec J}+\vec c_i\tilde{\vec C}_i+h\tilde H+d\tilde D+k\tilde K+m\tilde M.
\end{equation}
Again, $m$  is invariant under the coadjoint action; we assume that $m>0$. Consider the coadjoint action of $\exp(ix_k^aC_k^a)$. It reads
{\small
\begin{equation}
\label{e27}
\begin{split}
m'&=m,\\
j'^b&=j^b-\epsilon_{bad}\sum_{j=0}^Nx_j^ac_j^d-\frac m2\sum_{j=0}^N(-1)^{j-\frac{N+1}{2}}\epsilon_{bca}x_j^ax^c_{N-j}j!(N-j)!,\\
c'^b_j&=c_j^b+(-1)^{j-\frac{N-1}{2}}mj!(N-j)!x_{N-j}^b,\\
h'&=h+\sum_{j=0}^{N-1}(j+1)x_{j+1}^bc_j^b+\frac m2\sum_{j=1}^N(-1)^{j-\frac{N+1}{2}}j!(N-j+1)!x^a_jx^a_{N-j+1},\\
d'&=d-\sum_{j=0}^{N}(\frac N2-j)x_j^bc_j^b+\frac m2\sum_{j=0}^N(\frac N2-j)(-1)^{j-\frac{N+1}{2}}j!(N-j)!x^a_jx^a_{N-j},\\
k'&=k-\sum_{j=1}^{N}(N-j+1)x_{j-1}^bc_j^b+\frac m2\sum_{j=0}^{N-1}(-1)^{j-\frac{N-1}{2}}(j+1)!(N-j)!x^a_jx^a_{N-j-1},\\
\end{split}
\end{equation}
}
We see that, as in the case of Schr\"odinger group, any orbit contains the points 
\begin{equation}
\label{e28}
\vec s\tilde{\vec J}+(\chi^0-\chi^1)\tilde H+\chi^2\tilde D+(\chi^0+\chi^1)\tilde K+m\tilde M,
\end{equation}
where, again, $\vec s\in S^2$ and $\chi^\mu $ belongs to one of the orbits (\ref{e10}). The whole orbit is produced by acting with $\exp(ix_k^aC_k^a)$ on the above points. As a result we arrive at the following parametrization
\begin{align}
\label{e29}
j^b&=s^b-\frac m2\sum_{j=0}^N(-1)^{j-\frac{N+1}{2}}\epsilon_{bca}x_j^ax^c_{N-j}j!(N-j)!,\nonumber\\
c^b_j&=(-1)^{j-\frac{N-1}{2}}mj!(N-j)!x_{N-j}^b,\nonumber\\
h&=\chi^0-\chi^1+\frac m2\sum_{j=1}^N(-1)^{j-\frac{N+1}{2}}j!(N-j+1)!x^a_jx^a_{N-j+1},\\
d&=\chi^2+\frac m2\sum_{j=0}^N(\frac N2-j)(-1)^{j-\frac{N+1}{2}}j!(N-j)!x^a_jx^a_{N-j},\nonumber\\
k&=\chi^0+\chi^1+\frac m2\sum_{j=0}^{N-1}(-1)^{j-\frac{N-1}{2}}(j+1)!(N-j)!x^a_jx^a_{N-j-1}.\nonumber
\end{align}
The invariants $\vec s^2$ and $g_{\mu\nu}\chi^\mu\chi^\nu$, which characterize the orbits, correspond to the Casimir operators 
\begin{align}
\label{e30}
C_1&=M,\nonumber\\
C_2&=\left(M\vec J-\frac 12\sum_{j=0}^N\frac{(-1)^{j-\frac{N+1}{2}}}{j!(N-j)!}\vec C_j\times\vec C_{N-j}\right)^2,\\
C_3&=(MH-A)(MK-B)+(MK-B)(MH-A)-2(MD-C)^2,\nonumber
\end{align}
where
\begin{align}
\label{e31}
A&=\frac 12\sum_{j=1}^N\frac{(-1)^{j-\frac{N+1}{2}}}{(j-1)!(N-j)!}\vec C_{j-1} \vec C_{N-j},\nonumber\\
B&=-\frac 12\sum_{j=0}^{N-1}\frac{(-1)^{j-\frac{N+1}{2}}}{j!(N-j-1)!}\vec C_{j+1} \vec C_{N-j},\\
C&=\frac 12\sum_{j=0}^N\frac{(-1)^{j-\frac{N+1}{2}}}{j!(N-j)!}(j-\frac N2)\vec C_{j} \vec C_{N-j}.\nonumber
\end{align}
The basic dynamical variables are $\chi^\mu,s^a$ and $x_j$. The Poisson bracket resulting from Kirillov symplectic structure reads
\begin{equation}
\label{e32}
\{c_j^a,c_k^b\}=\delta^{ab}\delta^{N,j+k}(-1)^\frac{k-j+1}{2}k!j!m,
\end{equation}
and implies 
\begin{equation}
\label{e33}
\{x_k^a,x_{N-k}^b\}=\frac{\delta^{ab}(-1)^{k-\frac{N+1}{2}}}{mk!(N-k)!},\quad k=0,1,\ldots,N.
\end{equation}
It is easy to define Darboux coordinates for "external" variables. They read 
\begin{equation}
\label{e34}
\begin{split}
x_k^a &=\frac{(-1)^{k-\frac{N+1}{2}}}{k!}q_k^a,\\
x_{N-k}^a&=\frac{1}{m(N-k)!}p_k^a,
\end{split}
\end{equation}
for $\quad k=0,\ldots,\frac{N-1}{2}$, 
yielding the standard form of Poisson brackets
\begin{equation}
\label{e35}
\{q_k^a,p_l^b\}=\delta^{ab}\delta_{kl}.
\end{equation}
In terms of new variables the remaining one read
\begin{equation}
\label{e36}
\begin{split}
h&=\chi^0-\chi^1+\frac{1}{2m}{\vec p_{\frac{N-1}{2}}\vec p_{\frac{N-1}{2}}}+\sum_{k=1}^{\frac{N-1}{2}}\vec q_k\vec p_{k-1},\\
d&=\chi^2+\sum_{k=0}^{\frac{N-1}{2}}(\frac N2 -k)\vec q_k \vec p_k,\\
k&=\chi^0+\chi^1+\frac{m}{2}\left(\frac{N+1}{2}\right)^2{\vec q_{\frac{N-1}{2}}\vec q_{\frac{N-1}{2}}}-\sum_{k=0}^{\frac{N-3}{2}}(N-k)(k+1)\vec q_k\vec p_{k+1},\\
\vec j&=\vec s+\sum_{k=0}^{\frac{N-1}{2}}\vec q_k\times\vec p_k.
\end{split}
\end{equation}
The above findings can be compared with those of Ref. \cite{b29}.
In particular, the Hamiltonian $h$ is the sum of two terms depending on "internal" ($sl(2,\mR)$)  and "external" variables. The external part coincides with the Ostrogradski Hamiltonian \cite{b39} corresponding to the Lagrangian 
\begin{equation}
L=\frac{m}{2}\left(\frac{d^{\frac{N+1}{2}}\vec q}{dt^{\frac{N+1}{2}}}\right)^2.
\end{equation}
This can be easily seen by writing out the canonical equations of motion
\begin{equation}
\begin{split}
&\dot{\vec q}_k=\vec{q}_{k+1},\quad k=0,\ldots,\frac{N-3}{2},\\
&\dot{\vec p}_k=-{\vec{ p}}_{k-1},\quad k=1,\ldots,\frac{N-1}{2}\\
&\dot{\vec q}_{\frac{N-1}{2}}=\frac{1}{m}\vec p_{\frac{N-1}{2}},\quad \dot{\vec p}_0 =0
\end{split}
\end{equation}
which, for the basic variable $\vec q=\vec q_0$, imply ${\vec q}^{(N+1)}=0$.
\subsection{N-even}
As we have mentioned, in the case of  dimension $2$ for even $N$, there  exists also the  central extension of the Abelian algebra spanned by $\vec C$'s. The relevant commutators read:
\begin{equation}
\label{e37}
[C_j^a,C_k^b]=-i\epsilon^{ab}\delta^{N,j+k}(-1)^{\frac{j-k}{2}}k!j!M,
\end{equation}
where $\quad a,b=1,2,\quad j,k=0,1,\ldots,N$. Let us take an arbitrary element $X$ of dual space to the Lie algebra
\begin{equation}
\label{e38}
X=j\tilde{ J}+\vec c_i\tilde{\vec C}_i+h\tilde H+d\tilde D+k\tilde K+m\tilde M.
\end{equation}
As previously, $m$  is invariant under the coadjoint action; we can assume that $m>0$. Consider the coadjoint action of $\exp(ix_k^aC_k^a)$. It reads
\begin{equation}
\label{e39}
\begin{split}
m'=&m,\\
j'=&j-\epsilon^{ba}\sum_{j=0}^Nx_j^bc_j^a+\frac m2\sum_{j=0}^N(-1)^{\frac{2j-N}{2}}\epsilon^{ad}\epsilon^{bd}x_j^bx^a_{N-j}j!(N-j)!,\\
c'^b_j=&c_j^b-(-1)^{\frac{N-2j}{2}}mj!(N-j)!\epsilon^{ab}x_{N-j}^a,\\
h'=&h+\sum_{j=0}^{N-1}(j+1)x_{j+1}^bc_j^b+\frac m2\sum_{j=1}^N(-1)^{\frac{2j-N}{2}}j!(N-j+1)!\epsilon^{ab}x^b_jx^a_{N-j+1},\\
d'=&d-\sum_{j=0}^{N}(\frac N2-j)x_j^bc_j^b-\frac m2\sum_{j=0}^N(-\frac N2+j)(-1)^{\frac{2j-N}{2}}j!(N-j)!\epsilon^{ab}x^b_jx^a_{N-j},\\
k'=&k-\sum_{j=0}^{N-1}(N-j)x_{j}^bc_{j+1}^b-\frac m2\sum_{j=0}^{N-1}(-1)^{\frac{2j-N}{2}}(j+1)!(N-j)!\epsilon^{ba}x^b_jx^a_{N-j-1}.\\
\end{split}
\end{equation}
We see that, similarly to the case of $N$-odd, any orbit contains the points 
\begin{equation}
\label{e40}
s\tilde{ J}+(\chi^0-\chi^1)\tilde H+\chi^2\tilde D+(\chi^0+\chi^1)\tilde K+m\tilde M,
\end{equation}
where  $s\in \mR$ and $\chi^\mu $ belongs to one of the orbits (\ref{e10}).
Moreover, the whole orbit is produced by acting with $\exp(ix_k^aC_k^a)$ on the above points. Consequently, we have  the following parametrization
\begin{align}
\label{e41}
j&=s+\frac m2\sum_{j=0}^N(-1)^{\frac{2j-N}{2}}\epsilon^{ad}\epsilon^{bd}x_j^bx^a_{N-j}j!(N-j)!,\nonumber\\
c^b_j&=(-1)^{\frac{N-2j}{2}}mj!(N-j)!\epsilon^{ba}x_{N-j}^a,\nonumber\\
h&=\chi^0-\chi^1+\frac m2\sum_{j=1}^N(-1)^{\frac{2j-N}{2}}j!(N-j+1)!\epsilon^{ab}x^b_jx^a_{N-j+1},\\
d&=\chi^2-\frac m2\sum_{j=0}^N(-\frac N2+j)(-1)^{\frac{2j-N}{2}}j!(N-j)!\epsilon^{ab}x^b_jx^a_{N-j},\nonumber\\
k&=\chi^0+\chi^1-\frac m2\sum_{j=0}^{N-1}(-1)^{\frac{2j-N}{2}}(j+1)!(N-j)!\epsilon^{ab}x^b_jx^a_{N-j-1}.\nonumber
\end{align}
By direct, but rather tedious, computations we check that the  corresponding Casimir  operators are of the form
\begin{align}
\label{e42}
C_1&=M,\nonumber\\
C_2&=M J-\frac 12\sum_{j=0}^N\frac{(-1)^{\frac{2j-N}{2}}}{j!(N-j)!} C_{N-j}^aC^a_j,\\
C_3&=(MH-A)(MK-B)+(MK-B)(MH-A)-2(MD-C)^2,\nonumber
\end{align}
where
\begin{align}
\label{e43}
A&=\frac 12\sum_{j=1}^N\frac{(-1)^{\frac{2j-N}{2}}}{(j-1)!(N-j)!}\epsilon^{ab} C_{j-1}^b  C_{N-j}^a,\nonumber\\
B&=-\frac 12\sum_{j=0}^{N-1}\frac{(-1)^{\frac{2j-N}{2}}}{j!(N-j-1)!}\epsilon^{ab} C_{j+1}^b  C_{N-j}^a,\\
C&=\frac 12\sum_{j=0}^N\frac{(-1)^{\frac{2j-N}{2}}}{j!(N-j)!}(j-\frac{N}{2})\epsilon^{ab}C_{j}^b  C_{N-j}^a.\nonumber
\end{align}
The induced Poisson brackets of $\vec C$'s take the form 
\begin{equation}
\label{e44}
\{c_j^a,c_k^b\}=-\epsilon^{ab}\delta^{N,j+k}(-1)^\frac{k-j}{2}k!j!m,
\end{equation}
(for $\chi^\mu$ see eq. (\ref{e14})). Now let us define new coordinates as follows
\begin{equation}
\label{e45}
\begin{split}
x_j^a&=\frac{(-1)^{\frac{N-2j}{2}}}{j!}q_j^a, \quad j=0,\ldots,\frac{N}{2}-1,\quad a,b=1,2;\\
x_{N-j}^a&=\frac{1}{m(N-j)!}p_j^a,\quad j=0,\ldots,\frac{N}{2},\quad a,b=1,2.
\end{split}
\end{equation}
Then the nonvanishing Poisson brackets  read
\begin{equation}
\label{e46}
\begin{split}
\{q_j^a,p_j^b\}&=\delta^{ab}\delta_{jk},\quad j,k=0,\ldots ,\frac N2-1,\quad a,b=1,2;\\
\{q_{\frac N2}^a,q^b_{\frac N2}\}&=\frac 1m\epsilon^{ba},\quad a,b=1,2.
\end{split}
\end{equation}
Let us  introduce auxiliary notation (see eq. (32) in Ref. \cite{b29})
\begin{equation}
\label{e47}
p_{\frac N2}^a=\frac m2\epsilon ^{ba}q_{\frac N2}^b.
\end{equation}
Then, the remaining  dynamical variables take form
\begin{equation}
\label{e48}
\begin{split}
h&=\chi^0-\chi^1+\sum_{k=0}^{\frac{N}{2}-1}\vec p_k\vec q_{k+1},\\
d&=\chi^2+\sum_{k=0}^{\frac{N}{2}-1}(\frac N2 -k)\vec p_k \vec q_k,\\
k&=\chi^0+\chi^1-\sum_{k=1}^{\frac{N}{2}-1}(N-k+1)k\vec p_k\vec q_{k-1}-N(\frac N2+1)\vec q_{\frac N2 -1}\vec p_{\frac N2},\\
 j&=s+\sum_{k=0}^{\frac{N}{2}}\vec q_k\times \vec p_k.
\end{split}
\end{equation}
These results, in the case of trivial orbit $\mH_0$, agree with the ones obtained in Ref. \cite{b29}.
\par
We conclude that the general dynamical system admitting $N$-Galilean conformal symmetry with $N$-odd ($N$-even in dimension two) as the symmetry group acting transitively is described by the "external" variables corresponding to higher derivative Lagrangian and two kinds of internal ones: spin variables $\vec s$ (s, respectively) with trivial dynamics and $SL(2,\mR)$ spin variables $\chi^\mu$ with nontrivial conformal invariant one. As in the case of Schr\"odinger algebra it is easy to construct the explicitly time-dependent integrals of motion.
They generate the relevant symmetry transformations.
%%%%%%%%%%%%%%
\par
{\bf Acknowledgments} The authors would like to thank Professor Piotr Kosi\'nski for helpful discussions. We are grateful to prof. Peter Horvathy for useful remarks  which allowed us to improve the paper.
This work  is supported  in part by  MNiSzW grant No. N202331139


\begin{thebibliography}{99}
\bibitem{b1}\textsc{C.G.J. Jacobi}, Gesammelte Werke, Berlin Reimer (1884)
\bibitem{b2}\textsc{S. Lie}, Arch. Math. {\bfseries 6} (1881), 328  
\bibitem{b3}\textsc{R. Jackiw}, Phys. Today {\bfseries 25} (1972), 23
\bibitem{b4}\textsc{C.R. Hagen}, Phys. Rev.  {\bfseries D5} (1972), 377
\bibitem{b5}\textsc{U.Niederer}, Helv. Phys. Acta. {\bfseries 45} (1973), 802
\bibitem{b6}\textsc{C. Duval, G. Burdet, H.P. K\"unzle, M. Perrin}, Phys. Rev. {\bfseries D31} (1985), 1841
\bibitem{b7}\textsc{C. Duval, G.W. Gibbons, P.A. Horvathy}, Phys. Rev.  {\bfseries D43} (1991), 3907
\bibitem{b8}\textsc{M. Henkel}, J. Stat. Phys. {\bfseries 75} (1994), 1023
\bibitem{b9}\textsc{P. Havas, J. Pleba\'nski}, J. Math. Phys. {\bfseries 19} (1978), 482
\bibitem{b10}\textsc{C. Duval, P.A. Horvathy}, J. Phys.  {\bfseries A42} (2009), 465206
\bibitem{b11}\textsc{C. Duval}, Lecture Notes in Physics {\bfseries 261} 162, Springer 1986
\bibitem{b12}\textsc{C. Duval, M. Hassaine, P.A. Horvathy}, Ann. Phys. {\bfseries 324} (2009), 1158
\bibitem{b13}\textsc{C. Duval, S. Lazzarini}, arXiv:1201.0683
\bibitem{b15}\textsc{M. Blau, J. Hartong, B. Rollier}, JHEP {\bfseries 7} (2010), 069
\bibitem{b16}\textsc{J. Negro, M.A. del Olmo, A. Rodriguez-Marco}, J. Math. Phys. {\bfseries 38} (1997), 3786
\bibitem{b17}\textsc{J. Negro, M.A. del Olmo, A. Rodriguez-Marco}, J. Math. Phys. {\bfseries 38} (1997), 3810
\bibitem{b18}\textsc{J. Lukierski, P.C. Stichel, W.J. Zakrzewski}, Phys. Lett.  {\bfseries A357} (2006), 3810
\bibitem{b19}\textsc{J. Lukierski, P.C. Stichel, W.J. Zakrzewski}, Phys. Lett.  {\bfseries B650} (2007), 203
\bibitem{b20}\textsc{A.V. Galajinsky}, Phys. Rev.  {\bfseries D78} (2008), 087701
\bibitem{b21}\textsc{A. Bagchi, R. Gopakumar}, JHEP  {\bfseries 0907} (2009), 037
\bibitem{b22}\textsc{A. Bagchi, R. Gopakumar}, JHEP  {\bfseries 1008} (2010), 004
\bibitem{b23}\textsc{P.A. Horvathy, P.M. Zhang}, Eur. Phys. J.   {\bfseries C65} (2010), 607
\bibitem{b24}\textsc{C. Duval, P.A. Horvathy}, J. Phys. {\bf A44} (2011), 335203
\bibitem{b25}\textsc{A.V. Galajinsky, J. Masterov,} Phys. Lett.  {\bfseries B702} (2011), 265
\bibitem{b26}\textsc{S. Fedoruk, P. Kosi\'nski, J. Lukierski, P. Ma\'slanka},  Phys. Lett.   {\bfseries B699} (2011), 129  
\bibitem{b27}\textsc{S. Fedoruk, E. Ivanov, J. Lukierski}, Phys. Rev. {\bf D83} (2011), 085013
\bibitem{b28}\textsc{J. Lukierski}, arXiv:1101.4202
\bibitem{b29}\textsc{J. Gomis, K. Kamimura}, Phys. Rev. {\bf D85} (2012), 045023
\bibitem{b30}\textsc{ V. K. Dobrev, H.-D. Doebner, C. Mrugalla}, Rep. Math. Phys.  {\bfseries 39}  (1997),  201
\bibitem{b31}\textsc{ V. K. Dobrev, H.-D. Doebner, C. Mrugalla}, J. Phys.  {\bfseries A29}  (1996),  5909
\bibitem{b32}\textsc{ V. K. Dobrev, H.-D. Doebner, C. Mrugalla}, Mod. Phys. Lett.  {\bfseries A14}  (1999),  1113
\bibitem{b32b}\textsc{N. Aizawa, V.K. Dobrev}, Bucl. Phys.  {\bfseries B828}  (2010),  581
\bibitem{b32c}\textsc{M. Henkel, J. Unterberger}, Nucl. Phys.  {\bfseries B660}  (2003),  407
\bibitem{b32d}\textsc{M. Henkel}, Phys. Rev. Lett.  {\bfseries 78}  (1996),  1940
\bibitem{b32e}\textsc{R. Cherniha, M. Henkel}, J. Math. Anal. Appl.  {\bfseries 369}  (2010),  120
\bibitem{b33}\textsc{A. Kirillov}, Elements of the Theory of Representations, Springer 1976
\bibitem{b34}\textsc{J.M. Soriau}, Structure of Dynamical Systems. A Symplectic View of Physics, Birkhauser 1997
\bibitem{b35}\textsc{V.I. Arnold}, Mathematical Methods of Classical Mechanics, Springer 1989
\bibitem{b36}\textsc{R. Giachetti}, Riv. Nuovo. Cim. {\bf 4}, (1981), 1 
\bibitem{b37}\textsc{V. de Alfaro, S. Fubini, G. Furlan}, Nuovo Cim. {\bf A34} (1976), 569
\bibitem{b38}\textsc{K. Andrzejewski, J. Gonera}, arXiv:1108.1299
\bibitem{b14}\textsc{D. Martelli, Y. Tachikawa}, JHEP {\bfseries 05}, (2010), 091
\bibitem{b39}\textsc{M. Ostrogradski,} Mem. Acad. St. Petersburg {\bf 4} (1850), 385
\end{thebibliography}
\end{document}